\documentclass[letter]{aa}  
\usepackage{graphicx}
\usepackage{txfonts}
\usepackage{natbib}
\bibpunct{(}{)}{;}{a}{}{,} 
\usepackage{threeparttable}
\begin{document}
   \title{O and Na abundance patterns in open clusters of the Galactic disk}

   \subtitle{}

   \author{G. M. De Silva
          \inst{1}
          \and
          B. K. Gibson
          \inst{2}
           \and
          J. Lattanzio
           \inst{3}
          \and
          M. Asplund
           \inst{4}
          }

   \offprints{G.M. De Silva\\ \email{gdesilva@eso.org} }

   \institute{European Southern Observatory, Karl-Schwarzschild-Str 2, D-85748 Garching, Germany\
         \and
	 Jeremiah Horrocks Institute for Astrophysics \& Supercomputing, University of Central Lancashire, Preston, PR1 2HE, UK\
	\and
	   Centre for Stellar and Planetary Astrophysics, Monash University, Victoria 3800, Australia\
	\and
	   Max Planck Institute f\"{u}r Astrophysik, Postfach 1317, D-85741 Garching, Germany\
             }

   \date{Received ; accepted }

 
  \abstract
  {}
   {A global O-Na abundance anti-correlation is observed in globular clusters, which is not present in the Galactic field population. Open clusters are thought to be chemically homogeneous internally. We aim to explore the O and Na abundance pattern among the open cluster population of the Galactic disk.} 
   {We combine open cluster abundance ratios of O and Na from
  high-resolution spectroscopic studies in the literature and
  normalize them to a common solar scale. We compare the open
  cluster 
  abundances against the globular clusters and disk field.} 
   {We find that the different environments show different abundance
  patterns. The open clusters do not show the O-Na anti-correlation at
  the extreme O-depletion / Na-enhancement as observed in globular
  clusters. Furthermore, the high Na abundances in open clusters do not 
  match the disk field stars. If real, it
  may be suggesting that the dissolution of present-day open clusters
  is not a significant contribution to building the Galactic
  disk. Large-scale homogeneous studies of clusters and field 
  will further confirm the reality of the Na enhancement.} 
   {}


   \maketitle
%

\section{Introduction}
The O-Na abundance anti-correlation observed in Galactic globular
clusters (GCs) is thought to be an intrinsic property of the clusters
\citep[see review by][and references therein]{gratton04}. It is 
considered a global feature of GCs, because the anti-correlation has been
observed in all GCs subject to high-resolution studies
\citep{carretta06} and is present among the cluster giants, as well as
unevolved stars \citep{carretta04, ramirezcohen,
  gratton01}. Iron and the heavier elements show little star-to-star
abundance variations within a cluster \citep{suntzeff93}. The
currently supported theory for the origin of the O-Na anti-correlation
is that of primordial pollution from previous stars
\citep{cottrelldacosta}, although the 
mechanism responsible for the pollution is still
unclear. Several hypotheses are presently being
debated, including pollution by the ejecta of
massive stars \citep{decressin07a, decressin07b} or by  AGB stars undergoing
hot bottom burning \citep{ventura01, fenner04, karakas06}.

In general, the halo stars have a chemical composition similar to
that of the GCs with the exception of the lighter element abundance
trends \citep{gratton2000}. That the O-Na anti-correlation is not seen
in the halo stars presumably reflects the different chemical evolution
of the high-density cluster environment. Therefore stellar populations
showing 
GC-like chemical evolution can be isolated from other populations by
examining the O-Na abundance pattern \citep{geisler07}. The abundance
mismatch with the halo has important implications 
for the formation of the Galactic halo. It is
possible the halo was built up either from clusters that have long
since dissolved, which had a different chemical enrichment history to
the surviving GCs, or from the stars lost from the present-day GCs
before taking on the effects of the polluting stars.

A similar investigation can be applied to the progenitors of the
Galactic disk. It is often stated that open clusters (OCs) are the
objects of choice for tracing the star-formation history in the disk
\citep[e.g.][]{friel95}. Present-day OCs cover a wide range in 
age, metallicity, and position. They are used to examine the
disk chemical evolution, from the solar neighborhood to the outer
Galactic disk. Older OCs are considered particularly useful because they
provide a time line for change. In general the OC abundances match
that of the field, albeit with a larger scatter
\citep{friel02,desilva07,bragaglia08}. 

It is interesting that some OCs are Na-enhanced
\citep{friel05, jacobson07, bragaglia08}. Whether this is truly an intrinsic
property of the clusters or an artifact of the abundance measurements
remains unclear in the literature. A discrepancy in abundance 
patterns between the OCs and the disk has important implications for
our understanding of disk formation. Recall that the O and Na
abundance anomaly is the characteristic difference between the GCs and
the halo. To our knowledge no such characteristic abundance patterns have been
reported for OCs. In this Letter we explore the abundance patterns of
O and Na among the OC population and compare them against the GC abundance
anti-correlation and the disk field abundances. 

\section{Data samples}\label{sample}

We combine high-resolution studies that derive both O and Na
abundances in open clusters. The homogenized open cluster
  abundances, associated rms scatter, the number of stars per cluster,
  and references are given in Table \ref{tab1}. To represent the GC
  O-Na anti-correlation, we use abundances of red
giants in NGC~2808 as a template of the global GC O-Na
anti-correlation \citep[][see Fig. 5]{carretta06}. For comparisons
with the disk field, we use the compilation 
by \citet{soubirangirard} mainly based on field
dwarfs, the abundances published by \citet{mishenina06} based on field
red clump stars, and the disk giants studied by 
\citet{fulbright07}.

\begin{table}[h]
\begin{center}
\caption{Open cluster sample}\label{tab1}
\scriptsize
\begin{tabular}{lccccccccl}
\hline ID & [Fe/H] & $N$& [O/Fe]& $\sigma$& [Na/Fe]& $\sigma$& Ref. \\
\hline
Be 17	 &-0.10& 3& 0.13 &0.05 &0.15 & 0.08 &1\\
Be 20	 &-0.45& 2& 0.34 &0.01 &0.27 & 0.10 &2  \\
Be 29	 &-0.54& 2& 0.39 &0.03 &0.43 & 0.01 &2  \\
Be 31	 &-0.56& 1& 0.40 &  -  &0.34 &   -  &2  \\
Cr 261	 &-0.03& 6&-0.03 &0.09 &0.33 & 0.06 &3  \\
M 67	 & 0.03& 6&-0.07 &0.08 &-0.02& 0.07 &4\\
Mel 66	 &-0.38& 2& 0.45 &0.02 &0.24 & 0.03 &5  \\
Mel 71 	 &-0.30& 2&-0.20 &0.01 &0.2  & 0.05 &6 \\
NGC 2112 &-0.10& 2& 0.0  &0.04 &0.1  & 0.02 &6  \\
NGC 2141 &-0.18& 1& 0.16 & -   &0.48 &   -  &2  \\
NGC 2243 &-0.48& 2& 0.18 &0.09 &0.20 & 0.07 &5  \\
NGC 3680 &-0.04& 2& 0.20 &0.05 &0.01 & 0.08 &4\\
NGC 6253 &0.46 & 4&-0.12 &0.06 &0.21 & 0.02 &7 \\
NGC 6791 &0.47 & 4&-0.25 &0.08 &0.13 & 0.21 &7 \\
NGC 6939 &0.00 & 9& 0.04 &0.09 &0.32 & 0.11 &8 \\
NGC 7142 &0.14 & 4& 0.05 &0.03 &0.36 & 0.06 &8 \\
NGC 7789 &-0.04& 9&-0.07 &0.09 &0.28 & 0.07 &10 \\
Praesepe & 0.27& 7&-0.40 &0.20 &-0.04& 0.12 &4 \\
\hline
\end{tabular}
\end{center}
\scriptsize
References:
(1) \cite{friel05};         
(2) \cite{yong05}; 
(3) \cite{carretta05};      
(4) \cite{pace08};  
(5) \cite{gratton};      
(6) \cite{brown};      
(7) \cite{carretta07};      
(8) \cite{jacobson07};        
(9) \cite{jacobson08}; 
(10) \cite{tau05}.        
\end{table}

As always when comparing different literature sources, the presence
of systematic effects must be highlighted. As the systematics arise
at various stages, it is very cumbersome to accurately correct for
all possible effects. For example, the stellar effective
temperatures ($T_\mathrm{eff}$) could be subject to large systematic
differences. \citet{carretta07} and \citet{brown} derive
$T_\mathrm{eff}$ based on photometry, while other studies are based on
spectroscopy. A difference of 100K in $T_\mathrm{eff}$ may result in
Na abundance differences up to 0.1 dex. Since the studies are on
different clusters, such systematics are difficult to quantify. To
homogenize the different studies, we have normalized the published
abundances to the new solar values of log$\epsilon$ (O) = 8.70,
log$\epsilon$ (Na) = 6.21 and log$\epsilon$ (Fe) = 7.51
\citep{asplund09}. This normalization was applied to the abundances of
all OCs, GCs, and field stars, except for studies based on a strictly
differential analysis with respect to the Sun. 

Most of the studies use the same elemental lines
to derive oxygen and sodium abundances. For oxygen, either one or both
of the [O I] forbidden lines at 6300.3 \AA\ and 6363.8 \AA\ have been
used, where NLTE effects are negligible \citep{asplund05}. While all
other studies employ full spectral
synthesis to derive O abundances, \citet{carretta06} and
\citet{pace08} use EWs, taking the effects of line blending into account.

For sodium, most studies use Na I lines at 5682.6, 5688.2, 6154.2 and
6160.7 \AA. The $gf$ values used vary among the studies. In many
  cases this difference is small with negligible 
effects on abundances. The greatest differences are by \citet[][Be 17
  \& NGC 6939]{jacobson07}, where the adopted log $gf$ values are
higher by 0.2 - 0.3. As this results in a larger Na 
  abundance of about 0.2 - 0.3 dex, we adjust their published abundances to account for the offset in the $gf$ values. 

For the open cluster data, we use only published Na abundances based on 
LTE analysis in order to minimize systematics. The Na lines are,
however, subject to NLTE effects, where the NLTE corrections are
dependent on the stellar parameters, [Fe/H], and the Na
abundance \citep[e.g.][]{mashonkina,takeda}. While the cluster sample
spans almost one dex in [Fe/H], we do 
not see an Na abundance trend with metallicity when assuming LTE. Such a trend would
be expected in more metal-poor stars where the NLTE corrections are
larger. For the globular cluster NGC 2808, the authors correct their Na abundances for
NLTE effects following \citet{gratton99}. The Na abundances for the
field clump stars by \citet{mishenina06} is based on NLTE analysis. In
all other cases we adopt LTE abundances.

\section{Results and discussion}

\begin{figure}
\centering
\includegraphics[scale=0.4]{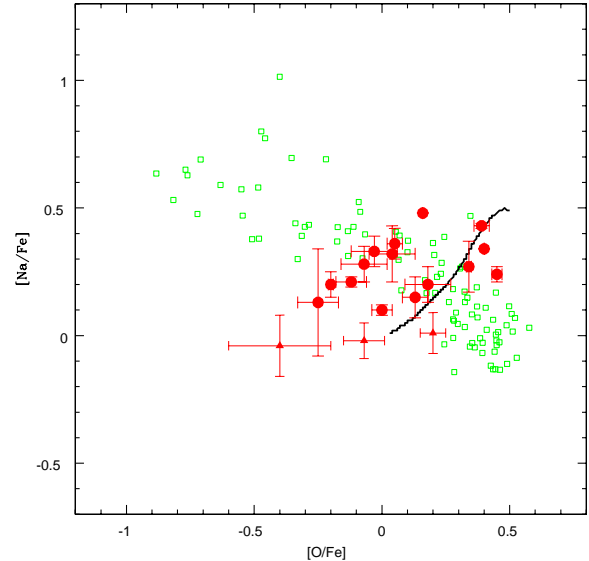}
\caption{Normalized [Na/Fe] vs. [O/Fe] for open clusters (red circles)
      compared to globular cluster NGC 2808
      (green squares). The red triangles are open clusters based on
      solar-type stars by \citet{pace08}. The black line represent a  
      Galactic chemical evolution model \citep{hughes}.}
\label{Fig1}
\end{figure}

In Figure \ref{Fig1} we plot the normalized [Na/Fe] and [O/Fe] for the
sample of OCs and GCs. Overplotted is the Galactic chemical evolution (GCE)
model that corresponds to the one described by \citet{hughes}. Only the 
final $\sim$4Gyrs (for [Fe/H] $>$ -1) is plotted for the annulus 
corresponding to the solar neighborhood. The plotted values for OCs
are the mean cluster abundances, whereas for the GCs the  
individual stars are plotted. We chose to plot only the mean value for
each open cluster because no significant star-to-star abundances
variations are found within an open cluster \citep{desilva06}. 

\begin{figure}
\centering
\includegraphics[scale=0.4]{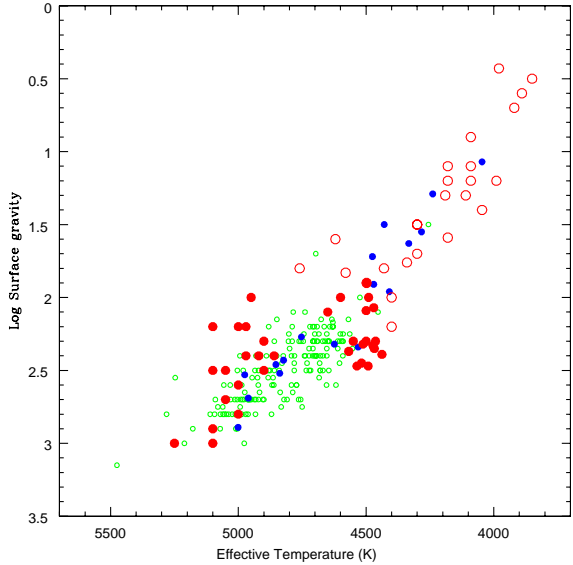}
\caption{Surface gravity vs. $T_\mathrm{eff}$ for open cluster stars compared to
  disk giants from \citet{mishenina06} (green) and
  \citet{fulbright07} (blue). Filled red circles are open cluster
  stars with $T_\mathrm{eff} >$ 4500 K and log g $>$ 2 and the open
  red circles are the remaining open cluster stars.}   
\label{Fig3}
\end{figure}

\begin{figure}
\centering
\includegraphics[scale=0.4]{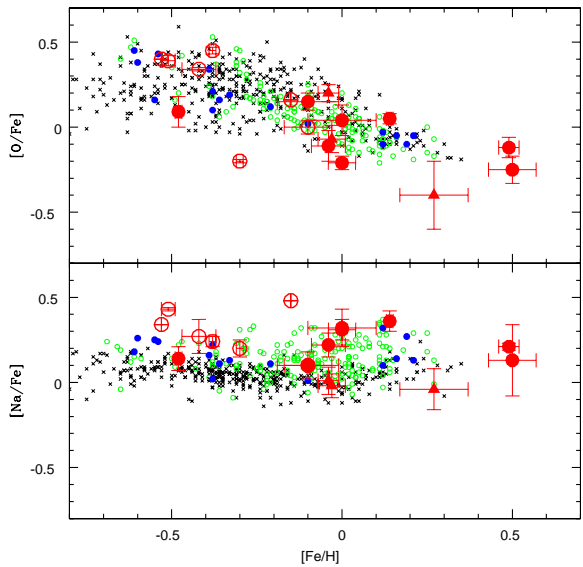}
\caption{[Na/Fe] and [O/Fe/] vs. [Fe/H] for open clusters and
  field. Red filled circles are cluster abundances from stars with
  $T_\mathrm{eff} >$ 4500 K and log g $>$ 2. Red triangles are clusters based on solar type stars by \citet{pace08}, and the red open circles are all other
  open clusters. Overplotted are disk dwarfs by \citet{soubirangirard}
  (crosses), disk clump stars by \citet{mishenina06} (green), and disk giants by \citet{fulbright07} (blue).} 
\label{Fig2}
\end{figure}

The OCs sit in the middle range of the O-Na anti-correlation of the
GCs. They do not reach the extreme values of O-depletion /
Na-enhancement and vice versa. If we disregard the only two clusters
with subsolar Na abundance (Praesepe and M67),
this might look as if the OCs follow the GC
anti-correlation, where all stars in a given OC have abundances
  similar to an individual star in a GC. The origins of the
  anti-correlation in GCs is thought to stem from primordial
  pollution. A plausible source is intermediate mass AGB stars undergoing hot bottom burning where the
temperatures are hot enough to activate the CNO and NeNa cycles
\citep{karakaslattanzio}. Here Ne is
used to produce Na via proton-capture, while O is depleted in the ON
cycle \citep{karakas06}. These AGB stars then pollute the proto-GC via
low-speed winds, which are retained in the cluster to form new stars
that are O-depleted and Na-enhanced \citep{ventura02, fenner04}.

Since all stars are chemically homogeneous in an OC, they must have
formed from the same (well mixed) gas cloud 
\citep{desilva06}; therefore, if pollution from an earlier generation
has taken place, it has to pollute all stars homogeneously or take
place before the cluster assembled. It is possible that a few
high-mass stars form shortly after the cloud assembles and enrich the
cloud fairly uniformly \citep{mckeetan02}. In such a scenario the
pollution effects are observed in the proceeding generation of stars
homogeneously within the cluster. 

Another possibility is that the first generation of OC stars were
dispersed and only the polluted stars are 
presently bound as a cluster. The level of pollution and the degree of
O-depletion and Na-enhancement of the polluting stars must vary at the
different site and time of the formation of the OCs in order
to account for the abundance spread across the OC sample.

The studies of \citet{bragaglia08} and \citet{desilva07} also
  suggest that Ba could be enhanced in 
OCs. The reality of this enhancement requires further
investigation, but it may indicate other sources responsible for pollution
in OCs. With the exception of NGC 1851 \citep{yonggrundahl}, abundance
variations of s-process elements in GCs are found to be insignificant
and unrelated to the O-Na anti-correlation. The likely source of
s-process elements are low mass AGB stars, whereas more massive AGB
stars are thought to be responsible for the O-Na anti-correlation in
GCs through hot bottom burning \citep{renzini}.

Figure \ref{Fig1} could be interpreted as the OC
population showing an O-Na abundance correlation, albeit with a wide
spread. The GCs also show a broad spread about the mean  
anti-correlation. Fitting a least square regression line provides a
correlation co-efficient of 0.5 when all OCs are considered, or
$\sim$0.78 when the data is split into two groups with a vertical
offset of 0.3 dex. The GCE model also shows an abundance correlation. The 
model corresponds to the one described by \citet{hughes}. A 'dual infall'
(halo + disk) framework for the solar neighborhood was  
employed in which the infall of primordial gas during the halo phase 
occurred on a rapid $\sim$50 Myr timescale; a second (disk) phase was
delayed by $\sim$1 Gyr with respect to the first, and then gas enriched
to 10\% solar meallicity was assumed to infall on an $\sim$10 Gyr
exponential timescale, thereafter. A conservative Schmidt Law for star
formation was assumed with an efficiency constrained by the
present-day gas fraction.  A \citet{kroupa93} IMF was employed with
lower and upper mass limits of 0.08 and 60M$_\odot$, respectively.
Most important, the adopted stellar yields include those of
\citet[][$>$10 M$_{\odot}$]{woosleyweaver95}, 
\citet[][$<$8 M$_\odot$]{karakaslattanzio}, and \citet[][Type Ia
  SNe]{nomoto}. We would expect a match in the abundance trend
between the OCs and the GCE model under the widely held assumption
that the OCs were a major contributor to building up the Galactic
disk. Slight variations in the model parameters would not be
sufficient to recover an O-Na abundance anti-correlation
\citep{fenner04}, and recovering such articulations would require
extreme conditions \citep{marcolini}.  

We now compare OCs against the disk field. To examine 
the stellar parameters of the different samples, Fig. \ref{Fig3}
plots the log $g$ vs. $T_\mathrm{eff}$ of the OC stars along with the
field stars by \citet{mishenina06} and \citet{fulbright07}. It shows
the open cluster stars with $T_\mathrm{eff} >$ 4500 K and log 
$g >$ 2, which match the \citet{mishenina06} stellar parameters. The
open circles show all other OC stars. Omitted from Fig.
\ref{Fig3} are the stars of \citet{soubirangirard}, which
are mostly dwarfs, and the OCs studied by \citet{pace08},
which are based on solar-type stars ($T_\mathrm{eff} >$ 5500 K and log $g >$
4). In Fig. \ref{Fig2} we plot [Na/Fe] and [O/Fe] vs. [Fe/H]
for the open clusters and disk field stars \citep[cf. Fig
  4. in][]{bragaglia08}.

The OC oxygen abundances match the various field samples, with the
exception of Praesepe and Mel 71, which are among the most 
O-depleted clusters. Sodium is significantly enhanced in OCs
compared to the field samples. As mentioned in Sect. \ref{sample},
we have only used OC LTE abundance. The \citet{soubirangirard} compilation also
uses LTE analyzes, but based on dwarf stars, whereas the OC stars
are mostly red giant and clump stars. It is possible 
that the OC giants are showing effects of internal mixing, making them
Na-enhanced compared to dwarfs. \citet{pasquini} 
find Na enhanced by 0.2 dex in giants compared to dwarf members
of OC IC 4561, while \citet{randich06} do not see abundance variations between main sequence and sub-giant stars in
M67. Therefore discrepancy with the \citet{soubirangirard} sample
could be due to the choice of targets.

The study by \citet{mishenina06} is of field clump stars, which match
some of the OC stellar parameters (see Fig. \ref{Fig3}); however,
they perform an NLTE analysis using the profiles and EWs of the Na
lines. They suggest the NLTE 
correction is about 0.1 to 0.15 where the LTE values are higher than
their NLTE equivalent. Further \citet{sestito08} find OC Na
abundances based on NLTE computations produce values lower by 0.1 - 0.2 dex
than using LTE assumptions. Therefore the
discrepancy between the OC and field clump stars may come from NLTE
effects. On the other hand, the abundances of disk giants by \citet{fulbright07}
are from an LTE analysis. Therefore the Na abundance mismatch between
the OCs and this sample is unclear, since here we are comparing similar
type stars (see Fig. \ref{Fig3}) and both are based on LTE analysis.

It is important to confirm this possible OC Na enhancement, because it has major
implications for the formation scenario of the Galactic disk. Likewise
with the GCs and the Galactic halo, it may imply that 
the OCs present today were not the major building blocks of
the Galactic disk. Rather, now dispersed clusters had a
different chemical evolution than those that have survived to date. The OC
abundances for other heavier elements match those of the disk field as is
the case between the GCs and the halo \citep{suntzeff93}. The
differences between the halo and GCs are highlighted mainly in the
lighter element abundances, notably the O-Na anti-correlation. 

The presented sample of OCs in this study are mostly old, over 1 Gyr
in age. Detailed abundance analysis of younger OCs are lacking in the
literature. It is expected that most clusters will dissolve into the
field within the first billion years \citep{lamers05}. The survival of
extremely old OCs around 10 Gyr in age is very interesting
as important fossils of the conditions of an earlier
era. One can question whether such surviving OCs are representative of the
chemical evolution of the disk, or rather whether embedded clusters
subject to high infant mortality rates are the major contributors to
the building of the Galactic disk \citep{lada2}. The older, still bound
clusters, which have deeper potentials, may in fact be the anomalous
objects that faced a different formation and chemical evolution. 

\section{Conclusions}

We have compiled literature studies of O and Na in OCs and
compared the abundance patterns with the GCs and disk
field stars. We homogenized the various studies to put them
  on the same abundance scale as best as possible. The OC population
  sits in the middle of 
the well-established GC anti-correlation, and it
is unclear whether they show an O-Na correlation or a GC-like
anti-correlation without reaching the extremes of O-depletion and
Na-enhancement.  

When compared to field abundances, the O abundances of OCs match the
field stars well. Na is enhanced
relative to the field. Much of the Na abundance discrepancy could be
explained by possible internal mixing in the giant stars and NLTE
effects; however, a possible intrinsic abundance mismatch cannot be
ruled out due to the mismatch between LTE open cluster giant
abundances and the LTE study of field giants by
\citet{fulbright07}. Homogeneous abundance studies of clusters and 
field is needed for confirmation. If the Na-enhancement is indeed a
true feature of old open clusters, it has major implications for 
understanding the formation of the Galactic disk. Further study of
O-poor clusters is encouraged to determine the 
O-Na abundance behavior at the extreme values of O-depletion.

\begin{acknowledgements}
We are grateful to the referee of this paper for useful suggestions. GMD thanks H. Lamers and M. Gieles for useful discussions and
T. Mishenina for providing information on the field analysis. BKG acknowledges the support of the UK's Science \& Technology Facilities Council (STFC Grant ST/F002432/1) and the Commonwealth Cosmology Initiative.
\end{acknowledgements}

\end{document}